# Subject Envelope based Multitype Reconstruction Algorithm of Speech Samples of Parkinson's Disease

Yongming Li[1*], Chengyu Liu[1], Pin Wang[1], Hehua Zhang[2], Anhai Wei[2]

(1. School of Microcommunication Engineering, Chongqing University, Chongqing 400044, P.R. China;
2. Department of Medical Engineering, Daping Hospital, Army Medical University (Third Military Medical University), Chongqing, 400038, China)

**Abstract**
The risk of Parkinson's disease (PD) is extremely serious, and PD speech recognition is an effective method of diagnosis nowadays. However, due to the influence of the disease stage, corpus, and other factors on data collection, the ability of every samples within one subject to reflect the status of PD vary. No samples are useless totally, and not samples are 100% perfect. This characteristic means that it is not suitable just to remove some samples or keep some samples. It is necessary to consider the sample transformation for obtaining high quality new samples. Unfortunately, existing PD speech recognition methods focus mainly on feature learning and classifier design rather than sample learning, and few methods consider the sample transformation. To solve the problem above, a PD speech sample transformation algorithm based on multitype reconstruction operators is proposed in this paper. The algorithm is divided into four major steps. Three types of reconstruction operators are designed in the algorithm: types A, B and C. Concerning the type A operator, the original dataset is directly reconstructed by designing a linear transformation to obtain the first dataset. The type B operator is designed for clustering and linear transformation of the dataset to obtain the second new dataset. The third operator, namely, the type C operator, reconstructs the dataset by clustering and convolution to obtain the third dataset. Finally, the base classifier is trained based on the three new datasets, and then the classification results are fused by decision weighting. In the experimental section, two representative PD speech datasets are used for verification. The results show that the proposed algorithm are effective. Compared with other algorithms, the proposed algorithm achieves apparent improvements in terms of classification accuracy.

## 1. INTRODUCTION

Parkinson's disease (PD) is a neurodegenerative disease. Currently, there is no permanent cure or prevention method for PD, but the disease can be controlled by early diagnosis and treatment. Dysarthria is an important early symptom of PD. Therefore, research on PD speech recognition algorithms based on machine learning is very important for diagnosing PD.

The relevant methods focus mainly on feature learning and classifier design. The main methods used for feature learning include PCA (principle component analysis, PCA) [1-3], LDA (linear discriminant analysis, LDA) [4,5], NNs (neural networks) [6-9], Relief [10], LPPs (locality preserving projections) [10], mRMR (maximum relevance and minimum redundancy) [11] and evolutionary computation [5,12-15]. The main classifiers include SVMs (support vector machines, SVMs) [16-19], KNN (K-nearest neighbor) classifiers [3,20], RFs (random forests, RFs) [16,21-23], ensemble learning classifiers [24-26] and decision trees [27]. Deep learning methods include DBNs (deep belief networks, DBNs) [28], DNNs (deep neural networks, DNNs) [29] and autoencoders [30]. Fuzzy theory [3,15,31-32] has also been adopted as an auxiliary method.

Although the methods above are conducive to improving classification accuracy, the quality problem of PD speech samples is still unsolved, so the improvement in accuracy is limited. The quality problem of PD speech samples is reflected mainly in the following aspects. Due to different associations between different corpora and PD dysarthrosis, the correlation between speech sample segments and disease category labels varies. For example, the major characteristic of the speech data of the PD are weak, low-pitched, metrical, and so on. The digit '3' can reflect the weak of speech data, but fail to reflect the rhythm of speech data. The short sentence and phrase with different unvoiced/voiced speech can reflect the metrical characteristic of the subject's speech data, but cannot reflect the breath of the subject well. Therefore, among these samples, no samples are useless, but no samples are useful with 100 percentages. So, it is not suitable to simply remove some samples or keep some samples.

For better description, the samples within one subject can be called sample segment. With different subjects, the sample segments are same. Sample segment selection can obtain high-quality samples from existing samples to improve the quality of samples. Methods of sample segment selection include sample selection of imbalanced datasets considering sparse neighborhoods [33], and the repeated clipping nearest neighbor method [34]. However, these methods are limited to the existing sample set and thus cannot reconstruct new samples. As the described above, since the sample segments are not satisfactory enough, the potential performance of these methods in improving sample quality is limited.

Therefore, it is worth to consider sample transformation to obtain new samples with higher quality. In recent years, some scholars have tried to study the transformation of PD speech samples to obtain new high-quality samples, thereby improving accuracy [35]. However, this method performs only linear reconstruction of all samples from a subject without considering other reconstruction methods and the difference among the sample segments. Therefore, the reconstructed PD speech sample segments are still unsatisfactory enough.

To solve the problem above, based the sample segments within same subject, this paper proposes a solution of sample transformation – a multitype reconstruction transformation algorithm for PD speech samples. The algorithm is divided into four major steps. First, a number of linear reconstruction operators are designed to transform the original data sample segments to obtain the first type of new sample segments. Second, considering the differences between sample segments, the original segments are clustered and reconstructed to obtain the second type of new segment. Third, the clustered sample segments are convolved with the linear reconstructed segments to obtain the third type of new segment. Finally, the base classifiers are trained based on the three types of new sample segments, and then the classification results are fused by decision level weighting. The segments are the samples of a subject and correspond to the corpus, which are the same for different subjects.

The main contributions and innovations of this paper are as follows.
1) For the first time, a clustering reconstruction operator of PD speech samples by clustering and linear transformation is proposed.
2) For the first time, a convolution reconstruction operator of PD speech samples achieved by clustering and convolution algorithm is proposed.
3) For the first time, using a multitype reconstruction transformation algorithm based on the three types of reconstruction operators designed is proposed to transform PD speech samples to obtain high-quality new samples, thereby improving classification accuracy apparently.

This paper is divided into four main sections. Section 1 describes the background, motivation and value of this paper. Section 2 describes the proposed method. Section 3 analyzes the experimental results. Section 4 discusses the contributions and potential value of this study and suggests future work.

## 2. Proposed Method - MRCST

### 2.1. Notation

| | |
|---|---|
| $S, \vec{S}_g, \widetilde{S}_l$ | original dataset, sample of $S$ 1…$G$, sample segment of subjects from $S$ 1…$L$ |
| $G$ | total number of original dataset samples |
| $N$ | number of features per sample |
| $L$ | number of subjects |
| $\ell(\cdot)$ | function of the refactoring operation |
| $\partial(\cdot)$ | function of the rounding operation |
| $\cdot$ | function of the dot multiplication |
| $Y$ | dataset after clustering $S$ |
| $\varphi(\cdot)$ | functions of the clustering |
| $\hat{Y}_l^q$ | sample segment of each cluster of $\vec{Y}_l$ 1…$Q$ |
| $Q$ | number of clustering centers |
| $\gamma(\cdot)$ | function of convolution operation |
| $[\vec{E}_1, \vec{E}_2, …, \vec{E}_6]$ | the mean, median, clipped mean (25%), standard deviation, quartile distance, mean absolute error of the sample set |
| $E_f, E_s, E_t$ | the target dataset of the first, second and third steps of the algorithm |

| $E'_f, E'_s, E'_t$ | normalized $E_f, E_s, E_t$ |
|---|---|
| $F_f, F_s, F_t$ | constructed $E'_f, E'_s, E'_t$ |
| $\Gamma(\cdot)$ | function of the matrix extension |

The original dataset $S = \begin{bmatrix} \vec{S}_1 \\ \vec{S}_2 \\ \cdots \\ \vec{S}_G \end{bmatrix} = \begin{bmatrix} s_{11} & s_{12} & \cdots & s_{1N} \\ s_{21} & s_{22} & \cdots & s_{2N} \\ \cdots & \cdots & \cdots & \cdots \\ s_{G1} & s_{G2} & \cdots & s_{GN} \end{bmatrix} = \begin{bmatrix} \widetilde{S}_1 \\ \widetilde{S}_2 \\ \cdots \\ \widetilde{S}_L \end{bmatrix}$, where $\vec{S}_i = [s_{i1}, s_{i2}, \ldots, s_{iN}], 1 \le i \le G$; the partitioned matrix $\widetilde{S}_i = \begin{bmatrix} s_{11} & s_{12} & \cdots & s_{1N} \\ s_{21} & s_{22} & \cdots & s_{2N} \\ \cdots & \cdots & \cdots & \cdots \\ s_{G_0 1} & s_{G_0 2} & \cdots & s_{G_0 N} \end{bmatrix} = \begin{bmatrix} \vec{S}_1 \\ \vec{S}_2 \\ \cdots \\ \vec{S}_{G_0} \end{bmatrix}, 1 \le i \le L$; $G$ is the total number of samples; $N$ is the number of features per sample (number of vector components); and all the samples belong to subject $L$; that is, $G_0 = G/L$ is the number of samples for each subject.

### 2.2. Type A reconstruction operator - direct linear reconstruction of the samples (EF_IT)

The sample segment $\widetilde{S}_i$ of subjects from the original dataset $S$ was reconstructed by Formulas (2) - (7) to obtain the target dataset $E_f = [E_{f1}, E_{f2}, \ldots, E_{fL}]^T$ as follows:

$$E_{fi} = \begin{bmatrix} \vec{E}_1 \\ \vec{E}_2 \\ \cdots \\ \vec{E}_6 \end{bmatrix} = \ell(\widetilde{S}_i) = \ell \left( \begin{bmatrix} \vec{S}_1 \\ \vec{S}_2 \\ \cdots \\ \vec{S}_{G_0} \end{bmatrix} \right), 1 \le i \le L \quad (1)$$

where $\ell(\cdot)$ is the reconstruction operator and the central tendency and dispersion index of the sample features are calculated by the reconstruction operator. These indexes are the mean, median, trimmed mean (25%), standard deviation, interquartile distance, and mean absolute error. The calculation formula for each index is as follows:

$$\vec{E}_1 = \frac{1}{G_0} \sum_{i=1}^{G_0} \vec{S}_i, \quad (2)$$

$$\vec{E}_2 = \begin{cases} \vec{S}_{(G_0+1)/2}, & G_0 \text{ is odd} \\ \frac{\vec{S}_{(G_0/2)} + \vec{S}_{(G_0/2+1)}}{2}, & G_0 \text{ not odd} \end{cases} \quad (3)$$

$$\vec{E}_3 = \frac{1}{G_0} \sum_{i=\partial(0.25 \times G_0)}^{G_0 - \partial(0.25 \times G_0)} \vec{S}_i \quad (4)$$

where $\partial(\cdot)$ is the rounding operation.

$$\vec{E}_4 = \sqrt{\frac{\sum_{i=1}^{G_0} (\vec{S}_i - \vec{E}_1) \cdot (\vec{S}_i - \vec{E}_1)}{G_0 - 1}} \quad (5)$$

where the notation $\cdot$ denotes the dot multiplication.

$$\vec{E}_5 = \vec{S}_{\partial(0.75 \times G_0)} - \vec{S}_{\partial(0.25 \times G_0)} \quad (6)$$

$$\vec{E}_6 = \frac{1}{G_0} \sum_{i=1}^{G_0} |\vec{S}_i - \vec{E}_1| \quad (7)$$

### 2.3. Type B reconstruction operator - clustering reconstruction of the samples (ES_IT)

Iterative means clustering is used for the sample segment $\widetilde{S}_i$ of subjects from the original dataset $S$ to obtain the dataset $Y = [\widetilde{Y}_1, \widetilde{Y}_2, \ldots, \widetilde{Y}_L]^T$ by formulas (8) - (11) as follows:

$$\widetilde{Y}_i = \begin{bmatrix} \hat{Y}_i^1 \\ \hat{Y}_i^2 \\ \vdots \\ \hat{Y}_i^Q \end{bmatrix} = \varphi(\widetilde{S}_i) = \begin{bmatrix} \varphi_{part1}(\widetilde{S}_i) \\ \varphi_{part2}(\widetilde{S}_i) \\ \vdots \\ \varphi_{partQ}(\widetilde{S}_i) \end{bmatrix}, 1 \le i \le L \quad (8)$$

where $\varphi(\cdot)$ is the cluster computing, $\hat{Y}_i^j$ and $\varphi_{partj}(\widetilde{S}_i)$ are the $i$th cluster sample segments of the $j$th subject, and $Q$ is the number of clustering centers. The iterative mean clustering algorithm refers to the construction of a layer of new samples based on the k-means clustering algorithm; these samples are then used as input samples for k-means clustering to obtain the next layer of new samples, and the above process is repeated. The formula used by the algorithm to calculate the distance between samples is as follows:

$$d(\vec{S}_i, \vec{S}_j) = \sqrt{(\vec{S}_i - \vec{S}_j)^T (\vec{S}_i - \vec{S}_j)} \quad (9)$$

$\vec{S}_i$ and $\vec{S}_j$ are individual samples. The minimization objective function is called the sum of squares error (SSE), the formula for which is as follows:

$$SSE = \sum_{k=1}^{K} \sum_{\vec{S}_i \in c_k} \|\vec{S}_i - c_k\|^2 \quad (10)$$

where $K$ is the number of samples in the cluster and $c_k$ is the $k^{th}$ clustering center. The updating formula of the cluster center is as follows:

$$c_k = \frac{\sum_{\vec{S}_i \in c_k} \vec{S}_i}{K} \quad (11)$$

Then, the linear reconstruction operator is applied to the sample segment of the subject clustering cluster of dataset $Y$ to obtain the target data set $E_s = [E_{s1}, E_{s2}, \ldots, E_{sL}]^T$ as follows:

$$E_{si} = \ell(\widetilde{Y}_i) = \begin{bmatrix} \ell(\hat{Y}_i^1) \\ \ell(\hat{Y}_i^2) \\ \ldots \\ \ell(\hat{Y}_i^Q) \end{bmatrix}, 1 \leq i \leq L \quad (12)$$

## 2.4. Type C reconstruction operator - convolution reconstruction of the samples (ET_IT)

A convolution reconstruction operator is designed to convolve datasets $Y$ and $E_s$ to obtain target dataset $E_t = [E_{t1}, E_{t2}, \ldots, E_{tL}]^T$ as follows:

$$E_{ti} = \gamma(\widetilde{Y}_i, \ell(\widetilde{Y}_i)) = \gamma(\widetilde{Y}_i, E_{si}) = \begin{bmatrix} \gamma(\hat{Y}_i^1, \ell(\hat{Y}_i^1)) \\ \gamma(\hat{Y}_i^2, \ell(\hat{Y}_i^2)) \\ \ldots \\ \gamma(\hat{Y}_i^Q, \ell(\hat{Y}_i^Q)) \end{bmatrix} = \begin{bmatrix} E_{ti}^1 \\ E_{ti}^2 \\ \ldots \\ E_{ti}^Q \end{bmatrix}, 1 \leq i \leq L \quad (13)$$

where $E_{ti}^q = \gamma(\hat{Y}_i^q, \ell(\hat{Y}_i^q)), 1 \leq q \leq Q$. $\gamma(\cdot)$ is the convolution operator. For sample sets $X$ and $T$, convolution formula (16) is used for $T$ and $X$ to obtain sample set $E_{ti}^q = \gamma(X, T) = \gamma(\hat{Y}_i^q, \ell(\hat{Y}_i^q)) = [\vec{E}_1', \vec{E}_2', \ldots, \vec{E}_M']^T$.

$$X = \begin{bmatrix} x_{11} & x_{12} & \ldots & x_{1N} \\ x_{21} & x_{22} & \ldots & x_{2N} \\ \ldots & \ldots & \ldots & \ldots \\ x_{I1} & x_{I2} & \ldots & x_{IN} \end{bmatrix} = \begin{bmatrix} \vec{X}_1 \\ \vec{X}_2 \\ \ldots \\ \vec{X}_I \end{bmatrix} = [x_1 \quad x_2 \quad \ldots \quad x_N], \quad T = \begin{bmatrix} t_{11} & t_{12} & \ldots & t_{1N} \\ t_{21} & t_{22} & \ldots & t_{2N} \\ \ldots & \ldots & \ldots & \ldots \\ t_{M1} & t_{M2} & \ldots & t_{MN} \end{bmatrix} = \begin{bmatrix} \vec{T}_1 \\ \vec{T}_2 \\ \ldots \\ \vec{T}_M \end{bmatrix},$$

where $X = \hat{Y}_i^q$, $T = \ell(\hat{Y}_i^q)$, $x_i = [x_{1i} \quad x_{2i} \quad \ldots \quad x_{Ii}]^T$ $I$ and $M$ are the number of samples. The convolution formula is as follows:

$$\vec{E}_{m,n}' = T_{m,n} * X_{\#,n} = \sum_{i=1}^{L} t_{m,n} \cdot x_{i,n} \quad (14)$$

where $T_{m,n} = t_{m,n}$, $X_{\#,n} = x_n$ is the $n$th column element of matrix $X$ and the notation $*$ denotes a two-dimensional convolution.

$$\vec{E}_m' = \sum_{i=1}^{I} \vec{T}_m \cdot \vec{X}_i, 1 \leq m \leq M \quad (15)$$

Formula (15) can be re-expressed as follows:

$$\vec{E}_m' = x' u_m \quad (16)$$

where $x' = \begin{bmatrix} x_1^T & x_2^T & \cdots & x_N^T \end{bmatrix}$, $u_m = \begin{bmatrix} t_1 & & \\ & \ddots & \\ & & t_N \end{bmatrix}$ consists of diagonally arranged column elements of the extended matrix

$\Gamma(\vec{T}_m)$. $\Gamma(\vec{T}_m) = \begin{bmatrix} T_1' \\ T_2' \\ \cdots \\ T_I' \end{bmatrix} = \begin{bmatrix} t_1 & t_2 & \cdots & t_N \end{bmatrix}$ extends the $1 \times N$ matrix $\vec{T}_m$ by copying it into an $I \times N$ matrix, where

$T_i' = \vec{T}_m, 1 \leq i \leq I$. With further derivation, the formula can be obtained as follows:

$$\gamma(X, T) = (x'U)^T \qquad (17)$$

where $U = \begin{bmatrix} u_1 & u_2 & \cdots & u_m \end{bmatrix}$.

*2.5. Weighted decision fusion of multiple-type reconstruction operators*

Datasets $E_f$, $E_s$ and $E_t$ are normalized to obtain datasets $E_f'$, $E_s'$ and $E_t'$ and to divide $E_f' = \begin{bmatrix} E_{ftrain} \\ E_{ftest} \end{bmatrix}$, $E_s' = \begin{bmatrix} E_{strain} \\ E_{stest} \end{bmatrix}$ and $E_t' = \begin{bmatrix} E_{ttrain} \\ E_{ttest} \end{bmatrix}$ into a training set and test set. Based on each reconstructed sample set, new training and test sample sets $F_f = \begin{bmatrix} F_{E_{ftrain}} \\ F_{E_{ftest}} \end{bmatrix}$, $F_s = \begin{bmatrix} F_{E_{strain}} \\ F_{E_{stest}} \end{bmatrix}$ and $F_t = \begin{bmatrix} F_{E_{ttrain}} \\ F_{E_{ttest}} \end{bmatrix}$ are constructed. Based on the three datasets $F_f$, $F_s$ and $F_t$, each subclassifier is trained, and the classification results are fused by decision weighting. The formula for the weighted fusion is as follows:

$$H_{final} = \sum_{i=1}^{3} \alpha_i H_i$$
$$s.t. 1 = \sum_{i=1}^{3} \alpha_i \qquad (18)$$

where $H_i$ is the prediction result of the $i$th classifier and $\alpha_i$ is the weight of the $i$th classifier, which is obtained by the grid search method.

*2.6. Multitype reconstruction and clustering transformation ensemble algorithm for PD Speech data (MRCST)*

The main flow of this proposed algorithm is divided into four steps. First, the linear reconstruction operator is designed to transform the data sample segments of each subject to obtain the first new dataset (EF_IT). Second, considering the differences between samples, the clustering reconstruction operator is designed to cluster the data sample segments of each subject and perform linear reconstruction of each cluster of the sample segments to obtain a second new data set (ES_IT). Third, the convolution reconstruction operator is designed to convolve the new sample set generated by clustering reconstruction with the clustered samples to obtain the third new dataset. Finally, the three new datasets are trained consecutively on base classifiers. Then, the test results of each classifier are fused. The main flow of this proposed algorithm can be seen in the pseudo code as follows.

---

*Algorithm: Multitype Reconstruction and Clustering Transformation Algorithm for PD Speech Data (MRCST)*

**Input:** Public dataset $S$, total sample size $G$, number of features per sample $N$, number of subjects
**Output:** Accuracy, Sensitivity, Specificity
**Procedure:**

1: Using Formulas (1)-(7), the linear reconstruction $\ell(\cdot)$ is used for the public dataset $S = \begin{bmatrix} \tilde{S}_1 \\ \tilde{S}_2 \\ \cdots \\ \tilde{S}_L \end{bmatrix}$ to obtain dataset

$E_f = \begin{bmatrix} \ell(\tilde{S}_1) \\ \ell(\tilde{S}_2) \\ \cdots \\ \ell(\tilde{S}_L) \end{bmatrix}$;

2: Using Formulas (8)-(12), clustering algorithm $\varphi(\cdot)$ is used for public dataset $S$ to obtain dataset

$$Y = \begin{bmatrix} \varphi(\tilde{S}_1) \\ \varphi(\tilde{S}_2) \\ \dots \\ \varphi(\tilde{S}_L) \end{bmatrix} = \begin{bmatrix} \tilde{Y}_1 \\ \tilde{Y}_2 \\ \dots \\ \tilde{Y}_L \end{bmatrix}$$, and linear reconstruction is used for dataset $Y$ to obtain dataset $E_s = \begin{bmatrix} \ell(\tilde{Y}_1) \\ \ell(\tilde{Y}_2) \\ \dots \\ \ell(\tilde{Y}_L) \end{bmatrix}$ according to Formulas (1) - (6);

3: Using Formulas (14)-(17), the convolution operator is used for datasets $Y$ and $E_s$ to obtain dataset

$$E_t = \begin{bmatrix} \gamma(\tilde{Y}_1, \ell(\tilde{Y}_1)) \\ \gamma(\tilde{Y}_2, \ell(\tilde{Y}_2)) \\ \dots \\ \gamma(\tilde{Y}_L, \ell(\tilde{Y}_L)) \end{bmatrix};$$

4: Normalize $E_f, E_s, E_t$ to obtain $E'_f, E'_s, E'_t$ and construct dataset $F_f, F_s, F_t$;

5: Using Formula (18), based on datasets $F_f$, $F_s$ and $F_t$, each subclassifier is trained, consecutively, and the results of each classification are weighted and fused;

6: Based on the LOSO CV, the classification accuracy, sensitivity and specificity are calculated.

## 3. EXPERIMENTAL RESULTS AND ANALYSIS

### 3.1. Experimental environment

A.1 Dataset

The two representative public PD speech datasets are chosen for verifying the proposed algorithm. The datasets are derived from the machine learning database established by the University of California, Irvine: the 1st dataset was created by Sakar et al.[17] in 2014. In the Sakar database, there are 40 subjects, including 20 patients with Parkinson's disease (6 women and 14 men) and 20 healthy people (10 women and 10 men). Each subject contains 26 speech sample segments. For each speech sample, 26 features are extracted to form a feature vector. The second dataset was created by Little et al.[36]. The dataset is composed of a range of biomedical voice measurements from 31 people, 23 with Parkinson's disease (7 women and 16 men) and 8 healthy people (5 women and 3 men). Each subject contains 6 or 7 speech samples. For each speech sample, 22 features are extracted to form a feature vector. For more detailed information on the first and second datasets, please visit the website (https://archive.ics.uci.edu/ml/index.php). Table I shows the basic information of the two datasets.

TABLE I
BASIC INFORMATION ON THE DATASET

| Dataset | Attribute | | | | | | |
|---|---|---|---|---|---|---|---|
| | Patients | Healthy people | Instances | Number of samples per subject | Futures | Classes | Reference |
| Sakar | 20 | 20 | 1040 | 26 | 26 | 2 | [17] |
| MaxLittle | 23 | 8 | 189 | 6 or 7 | 22 | 2 | [36] |

A.2 Experimental condition

The leave-one-subject-out (LOSO) method is applied here since multiple samples correspond to one subject in the dataset. This verification method can maximize the number of training samples, especially in small sample cases, and thus can better reflect the potential of the classification algorithm. Moreover, all the samples are fully tested, so the test accuracy is closer to the results in practical application scenarios. The subjects in the dataset are subjected to cross-validation by the LOSO method. Some of the existing algorithms are based on k-fold and holdout cross-validation methods, and the training and test samples are possibly from the same subject, thereby leading to a classification accuracy that is unrealistic. Unlike the two algorithms, the LOSO can guarantee that training and test samples are from different subjects, thereby ensuring that the classification accuracy is realistic and consistent with the actual diagnosis.

The set of parameters in this paper is as follows: a polynomial kernel support vector machine (SVM) and random forest (RF) are used in the experiment. The loss function of the SVM is 10, the gamma function is 0.005, and the number of random forest classifiers is 50. The hardware configuration of the experimental platform is as follows: an Intel i5-3230 M CPU, 4 GB of memory and MATLAB R2017b.

A.3 Evaluation metrics

Three evaluation metrics, namely, accuracy, sensitivity, and specificity, were exploited to measure the performance of the proposed algorithm.

$$accuracy(ACC) = \frac{TP+TN}{TP+FP+TN+FN}$$

$$sensitivity(TPR) = \frac{TP}{TP+FN}$$

$$specificity(TNR) = \frac{TN}{FP+TN}$$

where $TP$ denotes the number of true positives, $FP$ denotes the number of false positives, $TN$ denotes the number of true negatives, and $FN$ denotes the number of false negatives.

### 3.2. Verifying the Main Parts of the Proposed Algorithm

In this experiment, the ablation method is used to verify the effectiveness of the proposed algorithm - MRCST. This MRCST algorithm is divided into four steps. Step 1: Linear reconstruction operators are used on the original dataset to obtain the first new dataset (EF_IT). Step 2: The cluster and reconstruction operator is conducted on the original dataset to obtain the second new dataset (ES_IT). Step 3: Convolution is conducted by combining ES_IT and the clustered samples, thereby obtaining the third new dataset (ET_IT). Finally, two classifiers are used based on these three data sets, and decision weighted fusion is performed to obtain the final classification results. The experimental results are shown in Tables II and III. An SVM (support vector machine) and RF (random forest) are adopted in the experimental section.

TABLE II
COMPARISON OF THE ABLATION METHODS FOR THE SAKAR DATASET

| Method | Base classifier | Accuracy (%) | Sensitivity (%) | Specificity (%) |
|---|---|---|---|---|
| Without sample transformation | SVM | 50 | 45 | 55 |
| | RF | 62.5 | 65 | 60 |
| EF_IT | SVM | 79.25 | 75 | 83.5 |
| | RF | 76.76 | 73 | 80.5 |
| ES_IT | SVM | 89 | 88.5 | **89.5** |
| | RF | 78 | 75.5 | 80.5 |
| ET_IT | SVM | 81.25 | 86.5 | 76 |
| | RF | 82.5 | 84.5 | 80.5 |
| MRCST | SVM | **89.5** | **92.5** | 86.5 |
| | RF | **86** | 89 | 83 |

TABLE III
COMPARISON OF THE ABLATION METHODS FOR THE MAXLITTLE DATASET

| Method | Base classifier | Accuracy (%) | Sensitivity (%) | Specificity (%) |
|---|---|---|---|---|
| Without sample transformation | SVM | 87.24 | 95.24 | 66.25 |
| | RF | 80.5 | 97.14 | 37.5 |
| EF_IT | SVM | 88.28 | 95.24 | 70 |
| | RF | 80.34 | 90.95 | 52.5 |
| ES_IT | SVM | **93.1** | **100** | **75** |
| | RF | 83.1 | 96.67 | 47.5 |
| ET_IT | SVM | 91.72 | 98.57 | 73.75 |
| | RF | 83.79 | 96.19 | 50.12 |
| MRCST | SVM | 91.38 | 99.52 | 70 |
| | RF | 85.86 | 96.67 | 57.5 |

As shown in Table II, for the Sakar dataset, the accuracies of the SVM and RF on the original samples are the lowest. The accuracies of the SVM and RF are only 79.25% and 76.76%, respectively, on dataset 1 (i.e., EF_IT). On dataset 2 (i.e., ES_IT), the accuracies of the SVM and RF increased by 9.75% and 1.24%, respectively. Compared with ES_IT, the accuracy of the RF classifier on dataset 3 (i.e., ET_IT) increased by 4.5%. For MRCST, whether in the SVM or RF, the accuracy is higher than on the three datasets above (datasets 1, 2 and 3). The results in Tables II and III are similar, thus indicating that the new sample set obtained by the proposed algorithm can significantly improve the classification accuracy. As shown in Table III, for the MaxLittle dataset, the accuracies of the SVM and RF are 88.28% and 80.34%, respectively, for dataset 1. On the MaxLittle dataset, considering the differences in sample distribution, the clustering reconstruction method is adopted, and the accuracies of the SVM and RF are improved by 4.82% and 2.76%, respectively. For MRCST, the accuracy of the RF is higher than that on the above three datasets. In conclusion, the three types of new samples constructed by the proposed algorithm have higher accuracy than the original dataset without transformation, thus indicating that the proposed algorithm is effective. Moreover, the weighted fusion results of the three datasets achieve better accuracy on the proposed algorithm, thus indicating that the weighted fusion in the proposed algorithm is effective. In general, based on the analysis above, the three reconstruction operators are effective. Moreover, the fusion of the three reconstruction operators by the proposed MRCST leads to the best accuracy.

### 3.3. Comparison with classical feature learning methods

To further verify the effectiveness of the proposed method, classical feature extraction algorithms are selected to compare the results before and after using the proposed method. The comparison results are shown in Table VI.

TABLE IV
THE COMPARISON OF THE RESULTS OBTAINED BY MAINSTREAM FEATURE LEARNING ALGORITHMS ON THE SAKAR AND MAXLITTLE DATASETS

| Method | Classifier | Whether the MRCST was used | MaxLittle | | | Sakar | | |
|---|---|---|---|---|---|---|---|---|
| | | | Accuracy (%) | Sensitivity (%) | Specificity (%) | Accuracy (%) | Sensitivity (%) | Specificity (%) |
| LPP | SVM | N | 91.72 | 95.24 | 82.5 | 60 | 50 | 70 |
| | | Y | **95.17** | **99.52** | **83.75** | **89.5** | **90.5** | **88.5** |
| | RF | N | 85.52 | 99.05 | 50 | 62.5 | 60 | 65 |
| | | Y | **91.72** | 98.57 | 73.75 | **86.25** | 87 | 85.5 |
| PCA | SVM | N | 91.03 | 95.24 | 80 | 55 | 60 | 50 |
| | | Y | **94.48** | **99.52** | **81.25** | **87.75** | **91** | **84.5** |
| | RF | N | 82.07 | 99.05 | 37.75 | 62.5 | 65 | 60 |
| | | Y | **88.62** | 98.57 | **62.5** | **84.5** | **87.5** | **81.5** |
| LDA | SVM | N | 91.03 | 95.71 | 78.75 | 62.5 | 60 | 65 |
| | | Y | **94.14** | **99.52** | **80** | **89.5** | **90.5** | **87.5** |
| | RF | N | 84.83 | 98.1 | 50 | 57.5 | 60 | 55 |
| | | Y | **94.83** | 97.62 | **87.5** | **91.25** | **91.5** | **91** |

As shown in Table IV, on the Sakar dataset, introducing the proposed algorithm improves the accuracy of the four feature learning algorithms by at least 20%, which is a significant improvement. On the MaxLittle dataset, due to the imbalance of positive and negative samples, TPR (sensitivity) and TNR (specificity) vary greatly among the four comparison methods, and the proposed algorithm is better in specificity. The accuracy of the RF combined with the Relief algorithm is increased by 5.17%. The accuracy of the SVM combined with the LPP is increased by 3.45%. The accuracy of the RF combined with PCA is increased by 6.55%. The accuracy of the RF combined with LDA is increased by 10%. In conclusion, in most cases, the proposed algorithm significantly improves the performance of existing feature reduction algorithms.

*3.4. Comparison with PD speech recognition algorithms*

The comparison results of the proposed algorithm with existing representative PD speech algorithms on the Sakar dataset are presented in Table V. The comparison results on the MaxLittle dataset are shown in Table VI.

TABLE V
COMPARISON OF THE RESULTS OBTAINED BY MAINSTREAM PD SPEECH RECOGNITION ALGORITHMS ON THE SAKAR DATASET

| Study | Method | Accuracy (%) | Sensitivity (%) | Specificity (%) |
|---|---|---|---|---|
| Sakar et al.[35] | KNN+SVM | 55 | 60 | 50 |
| Canturk and Karabiber [36] | 4 Feature Selection Methods+ 6 Classifiers | 57.5 | 54.28 | 80 |
| Behroozi and Sami[25] | Multiple classifier framework | 87.50 | 90.00 | 85.00 |
| Zhang et al.[21] | MENN+RF with MENN | 81.5 | 92.50 | 70.50 |
| Benba et al.[38] | HFCC+SVM | 87.5 | 90.00 | 85.00 |
| Li et al.[19] | Hybrid feature learning+SVM | 82.50 | 85.00 | 80.00 |
| Benba et al.[18] | MFCC+SVM | 82.5 | 80.0 | 85.0 |
| Proposed algorithm | MRCSST+SVM | 89.5 | 92.5 | 86.5 |
| Proposed algorithm | MRCSST+LDA+RF | **91.25** | **91.5** | **91** |

TABLE VI
COMPARISON OF THE RESULTS OBTAINED BY MAINSTREAM PD SPEECH RECOGNITION ALGORITHMS ON THE MAXLITTLE DATASET

| Study | Method | Accuracy (%) | Sensitivity (%) | Specificity (%) |
|---|---|---|---|---|
| Luukka [31] | Fuzzy entropy measures+similarity | 85.03 (Holdout) | — | — |
| Spadoto et al.[14] | PSO+OPF harmony search+OPF gravitational search+OPF | 84.01 (Holdout) | — | — |
| Kadam and Jadhav [29] | FESA-DNN | 93.84 | 95.23 | 90 |
| Proposed algorithm | MRCSST+SVM | 91.4 | 99.5 | 70 |

| | Proposed algorithm | MRCSST+LDA+RF | **94.83** | 97.62 | 87.5 |

As shown in Table VII, without feature reduction learning, the average accuracy of the proposed algorithm reaches 89.5%, which is better than the accuracies of most of the compared algorithms. For the proposed algorithm with feature reduction, the accuracy is further improved by 1.7% and is the highest. As shown in Table VIII, the average accuracy of the proposed algorithm without feature reduction reaches 91.4%, which is better than the accuracies of most of the compared algorithms; the average accuracy of the proposed algorithm with feature reduction reaches 94.83%, thereby achieving the highest accuracy.

## 4. DISCUSSION AND CONCLUSIONS

Research on PD speech recognition algorithms based on machine learning is very important for improving the diagnosis of PD. Due to the influence of the disease degree of the data collection object, the characteristic of the corpus, and other factors, the representation of the acquired data samples for the class label (the disease status) vary, thus affecting classification accuracy. This problem has become the main obstacle to improving PD speech recognition. These sample selection methods are limited to the existing sample set and thus cannot reconstruct new samples. Therefore, it is worth to consider sample transformation to obtain new samples with higher quality. Unfortunately, few of the existing PD speech recognition algorithms consider solving the problem.

To solve the problems above, this paper proposes a solution — a multitype reconstruction transformation algorithm for transforming PD speech data (MRCST) to improve the quality of the original sample segment. The algorithm is divided into four main steps. First, a number of linear reconstruction operators are designed to transform the original data sample segments to obtain the first new dataset. Second, considering the differences between samples, the original dataset is clustered and reconstructed to obtain the second new dataset. Third, the new sample segments generated by clustering and reconstruction are convolved with the clustered sample segments to obtain the third new dataset. Finally, the base classifiers are trained based on the three new datasets, and then the classification results are fused by decision-level weighting.

The main contributions and innovations of this paper are as follows:

1) A PD speech sample transformation algorithm based on multitype reconstruction and clustering is proposed to obtain high-quality new samples effectively, thus improving recognition accuracy. This study can be helpful for obtaining new PD speech sample segments with higher quality.

2) In considering the differences among samples, the sample set was clustered before the linear reconstruction operator was adopted, thereby obtaining clustering reconstruction of PD speech sample segments.

3) The clustering convolution operator is designed to construct a new dataset.

Currently, there are few public datasets for PD speech diagnosis. In this paper, two representative PD speech datasets are selected for verification. A groups of experiments is conducted. The main innovation parts of the proposed algorithm are verified. Representative PD speech recognition algorithms are chosen for comparison. The experimental results show that the proposed algorithm is effective. The main purpose of this study is to reconstruct new high-quality samples and to design the speech sample transformation framework. Therefore, although four feature reduction algorithms and two classifiers are used in this paper, the proposed algorithm can be easily applied to other feature reduction algorithms and classifiers.

Future work potentially includes the following: 1) further research, including deep learning research, on the combination of the proposed algorithm and other feature learning algorithms; and 2) further research on other classification algorithms, such as ensemble learning.

ACKNOWLEDGMENTS

We are grateful for the support of the National Natural Science Foundation of China NSFC (No. 61771080); Natural Science Foundation of Chongqing (cstc2020jcyj-msxmX0100, cstc2020jscx-gksb0010, and cstc2020jscx-msxm0369); Basic and Advanced Research Project in Chongqing (cstc2018jcyjA3022, cstc2020jscx-fyzx0212, cstc2020jscx-msxm0369, cstc2020jcyj-msxmX0523，and cstc2020jscx-gksb0010); Chongqing Social Science Planning Project (2018YBYY133); and the special project for improving scientific and technological innovation ability of Army Medical University (2019XLC3055).